\documentclass[showpacs,pra,floatfix,twocolumn]{revtex4}
\usepackage{graphicx, times}
\usepackage{amsmath}
\usepackage{verbatim}
\newcommand{\ba}{\begin{eqnarray}}
\newcommand{\be}{\begin{equation}}
\newcommand{\ea}{\end{eqnarray}}
\newcommand{\ee}{\end{equation}}

\begin{document}
\title{Entanglement entropy of two-species hard-core bosons in one dimension}
\author{Tae Yong Kim}
\author{Hang Gon Cho}
\author{Ji-Woo Lee}
\email{jwlee@mju.ac.kr}
\thanks{Fax: +82-31-335-7248}
\affiliation{Department of Physics, Myongji University, Yongin 17052, Korea}

\date{\today}

\begin{abstract}
We study the von Neumann entropy of a model for two-species hard-core bosons in one dimension.
In this model, the same-species bosons satisfy hard-core conditions, while the different-species bosons are allowed to occupy the same site with a local interaction $U$.
At half-filling, by Jordan-Wigner transformation, the model is exactly mapped to a fermionic Hubbard model.
The phase transition from superfluid $U=0$ to Mott insulator $U>0$ can be explained by simple one-band theory at half-filling.
We measure the von Neumann entanglement entropy of the ground states for the half-filled case and away from half-filling to understand quantum phase transitions.
To achieve this goal, we use a time-evolution-block decimation method with infinite-size matrix product state and also use a density matrix renormalization group with matrix product operators with large bond dimensions up to 300.
We found strong evidence that the local minimum point of the von Neumann entanglement entropy is the quantum critical point for finite-bond-dimension matrix product states.

\end{abstract}


\maketitle


\section{Introduction}

Entropy is one of the most important quantities in physics.
The second law of thermodynamics states that the entropy of an isolated system never decreases.
For pure quantum states with definite energy, the entropy of that quantum state depends on the degeneracy of those quantum states.
This idea corresponds to the quantum mechanical description of a microcanonical ensemble in statistical physics\cite{Huang}.
In quantum physics and quantum information science, an interesting quantity called ``entanglement entropy'' has attracted a lot of attention recently\cite{EisertRMP, PRA01, PRA02}.

Entanglement entropy (EE) is a measure of the amount of entanglement between a given part of a system and the rest of the system.
If one takes a many-body quantum state randomly, the entanglement entropy of a part of the system scales with the volume of the given part, which is called ``volume law''.
But the ground state of a quantum system has a peculiar property that the entanglement entropy of some part of the system scales with the area of that part\cite{EisertRMP}.
Some of the integrable models including bosonic harmonic chain, fermionic chain, and XY model are studied to understand ``area law''.  
Also, a conformal field theory \cite{Cardy} has shown that, for critical systems, the entanglement entropy scales with the correlation length $\xi$ as $\frac{c}{6} \log_2 (\xi/a)$ where $a$ is the lattice constant.

The recent development of representing quantum many-body states with tensor network states\cite{Vidal2003} allowed the direct measurement of the entanglement entropy with singular value decomposition\cite{Golub}.
Because an exact diagonalization of a Hamiltonian is severely restricted by the exponential growth of Hilbert space with the number of quantum basis, it is nontrivial to obtain the density matrix of a partial system.
The evolution of tensor network states by density matrix renormalization group (DMRG) with matrix product operators\cite{McCulloch} and time-evolution of block decimation\cite{Vidal} (TEBD) opened a new way of obtaining the entanglement entropy.

In this paper, we study the entanglement entropy of ground states of a model of two-species hard-core bosons in one dimension at zero temperature. 
Here, the hard-core condition is that the same-species bosons cannot occupy the same site, but different-species bosons can occupy the same site.
For the two species, we denote them as $a$ and $b$ boson.
Therefore the possible quantum states for a single site are $|0\rangle,  |a\rangle, |b \rangle, |(ab) \rangle$, where $|0\rangle$ denotes the vacuum state.
Here, $|(ab) \rangle$ state will have the interacting term $U$.
The locally interacting term $U$ between different species bosons at the same site can be either attractive and repulsive.

The model can be regarded as the bosonic version of the Hubbard model in one dimension.
Especially, at half-filling, the model has three quantum phases at zero temperature: a superfluid, the Mott insulator, and paired charge density wave.
When $U$ is positive, the system behaves like an insulator, which is called the Mott insulator. The system is also an insulator when $U$ is negative because the paired bosons hinder the movement of other pairs and the phase is called paired charge density wave.
Because the quantum phase transition occurs at $U=0$, it is very important to understand the properties of ground states near $U=0$ for the half-filling case.
It is also interesting to understand the properties of ground states for the off-half-filling case.

  This work focuses on the behavior of entanglement entropy in the two-species boson model in one dimension, which is hard to be obtained from Bethe ansatz. To obtain this quantity, we use infinite-size time evolution block decimation (iTEBD) with infinite matrix product states (iMPS) and infinite-size numerical renormalization group with the matrix product operators (MPO)\cite{Vidal, McCulloch}.
We found that the entanglement entropy is locally minimum as a function of $U$ for the quantum phase transition and the derivative of energy is zero.

This paper is organized as follows: in the following Section II, we present our model of interacting two-species hard-core bosons with local interaction $U$ for different species. 
Section III describes the methods used in this work: iTEBD and iDMRG with MPO.
Especially we observed a fast convergence of variational energies for iTEBD but a different converging behavior for entanglement entropy.
We suggest a method that overcomes this difficulty.
Section IV shows the ground-state energies we obtained by iTEBD and iDMRG and compares the values with the exact numbers at half-filling.
Section V shows the main result of our works, which is von Neumann half-chain entanglement entropy in half-filled and away-from-half-filled cases.
In Section VI, we discuss our results and future research topic.

\section{Model}

The model we study in this work is a model for two-species bosons which are distinguished by the indices $a$ and $b$ in a one-dimensional lattice with a periodic boundary condition.
Each species has a hard-core condition so that the possible number of each boson species in each site is restricted to 0 or 1.
If there is no interaction between $a$ and $b$, the system can be regarded as two separate bosonic systems.
We introduce a local interaction of $U$ between $a$ and $b$ boson.
In this sense, this model is a minimal interacting model of a two-species bosonic system.

The Hamiltonian is written as follows:
\begin{equation}
\begin{split}
H = & -t \sum_{
\langle i, j \rangle, \sigma =a, b } ( c^+_{i \sigma} c_{j\sigma}  + H.c. )
+ U \sum_{i} (n_{ia }- \frac{1}{2}) ( n_{ib} - \frac{1}{2} ) \\
& - \mu \sum_i (n_{ia} + n_{ib}),
\label{Hamiltonian}
\end{split}
\end{equation}
where
$c^+_{i \sigma} (c)$ is a boson creation (destruction) operator at site $i$ with a species index of $\sigma$ ($a$ or $b$),
and $n_{i \sigma} \equiv c^+_{i\sigma} c_{i\sigma}$.
Here $t$ is the hopping energy between nearest neighbors, and $U$ is on-site interaction energy between $a$ and $b$ when they occupy the same site $i$.
In the following, the energies are scaled as we set $t=1$.
The half-filled case is realized when $\mu=0$ so that the only variable parameter becomes $U$.
We will also investigate for the case of $\mu \neq 0$.

\section{time-evolution-block-decimation method and density matrix renormalization group method}

The infinite-size matrix product state is given by
\begin{equation}
|{\rm iMPS} \rangle  = \sum_{\{p \} } \sum_{\{\alpha\} }\Gamma^{p_1}_{\alpha_1, \alpha_2} \lambda^{[1] }_{\alpha_2} \Gamma^{p_2}_{\alpha_2, \alpha_1} \lambda^{[2]}_{\alpha_1}   | p_1 p_2 \rangle
\end{equation}
where $\Gamma$'s are tensors with a physical variable $p$ and indices of $\alpha$, and $\lambda$'s are the singular values connecting two neighboring sites.
The time-evolution block decimation method\cite{Vidal} (iTEBD) makes a matrix product state evolve with the repeating operation of a time evolution operator, $e^{-\epsilon H}$.
In application of this time-evolution operator, we decompose the operator as
\begin{equation}
e^{-\epsilon H} \sim e^{-\epsilon H_{12} } \times e^{- \epsilon H_{21} }
\end{equation}
Here, the operator $e^{- \epsilon H_{12} }$ acts on
\begin{equation}
e^{-\epsilon H_{12}} \lambda^{[2]}_{\alpha_1} \Gamma^{p_1}_{\alpha_1 \alpha_2} \lambda^{[1]}_{\alpha_2} \Gamma^{p_2}_{\alpha_2 \alpha_1} \lambda^{[2]}_{\alpha_1}
\end{equation}
This procedure updates $\Gamma^{p_1}, \Gamma^{p_2}$ and $\lambda^{[1]}$ by singular value decomposition.
Similarly, by applying $e^{- \epsilon H_{21} }$, it updates $\Gamma^{p_2}, \Gamma^{p_1}$, $\lambda^{[2]}$.

Suppose that a random state $|\Psi \rangle$ is a sum of all the eigenstates of a given Hamiltonian as

\begin{equation}
    | \Psi \rangle = \sum_{i=0}^\infty C_i | \Phi_i \rangle 
\end{equation}.
Then if we apply $e^{-\epsilon H} $, this state becomes
\begin{eqnarray}
   e^{-\epsilon H} | \Psi \rangle & = & \sum_{i=0}^\infty  e^{-\epsilon E_i} C_i | \Phi_i \rangle    \nonumber     \\
    & = & C_0 e^{-\epsilon E_0} | \Phi_0 \rangle \bigg( 1 + \sum_{i=1}^\infty  C_i e^{-\epsilon(E_i - E_0)} |\Phi_i \rangle \bigg). \nonumber \\
\end{eqnarray}
The states with energies larger than the ground-state energy quickly diminish with the consecutive operation of $e^{-\epsilon H}$ so that we can extract the ground state.

The infinite-size density matrix renormalization group (iDMRG) method is transforming a matrix product state to another matrix product state by singular value decomposition of the lowest eigenvalue state for the block Hamiltonian.
We adopt a matrix product operator (MPO) scheme\cite{McCulloch} in making the block Hamiltonian.
The MPO we use in our work is given by
\begin{equation}
W=
\begin{pmatrix}
I & a^{\dagger} & a & b^{\dagger} & b & V_{loc} \\
0 &   0          & 0 &           0 & 0 & -t a \\
0 &   0          & 0 &           0 & 0 & -t a^{\dagger} \\
0 &   0          & 0 &           0 & 0 & -t b \\
0 &   0          & 0 &           0 & 0 & -t b^{\dagger} \\
0 &   0          & 0 &           0 & 0 & I
\end{pmatrix}.
\end{equation}
where $V_{loc}= -\mu( n_b+n_a) + U(n_a - \frac{1}{2} I )(n_b-\frac{1}{2}I )$.
Here, all the matrix size is $4\times4$ because the size of possible quantum state at a local site is 4, {\it i.e.}, vacuum, $a$ boson only, $b$ boson only, and $a$ and $b$ bosons.
The basic algorithm of iDMRG is as follows.
First, the two-site iDMRG wave function is constructed from two tensors and
singular value matrix.
\begin{equation}
s^{[0]} A^{[1]} A^{[2]}
\end{equation}
Two-site block Hamiltonian is constructed by $L$, $W$, $W$, and $R$ tensors.
\begin{equation}
H_B = LWWR
\end{equation}
After finding the lowest eigenvalue state, we apply singular value decomposition for the eigenstate.
\begin{equation}
U s^{[1]} V^T
\end{equation}
Then the new wave function is
\begin{equation}
{A^{[1]}}' = (s^{[0]})^{-1} U s^{[1]}
\end{equation}
and
\begin{equation}
{A^{[2]}}' = V^T
\end{equation}
The new environment tensor $L'$ and $R'$ is updated as
\begin{equation}
L' = LW{A^{[1]}}' ({A^{[1]}}')^*
\end{equation}

\begin{equation}
R' = W {A^{{2}}}' ({A^{[2]}}')^* R
\end{equation}
Substituting primed tensors to unprimed ones, 
we iterate this procedure until the measurement of energy per site reaches a certain criterion.
The iDMRG method has an advantage of i) not using the parameter $\epsilon$ in iTEBD and ii) no need to construct a density matrix as in the old DMRG algorithm\cite{White}.

\section{On the convergence of energy and half-chain entanglement entropy}

\begin{figure}[t]
\includegraphics[width=8cm]{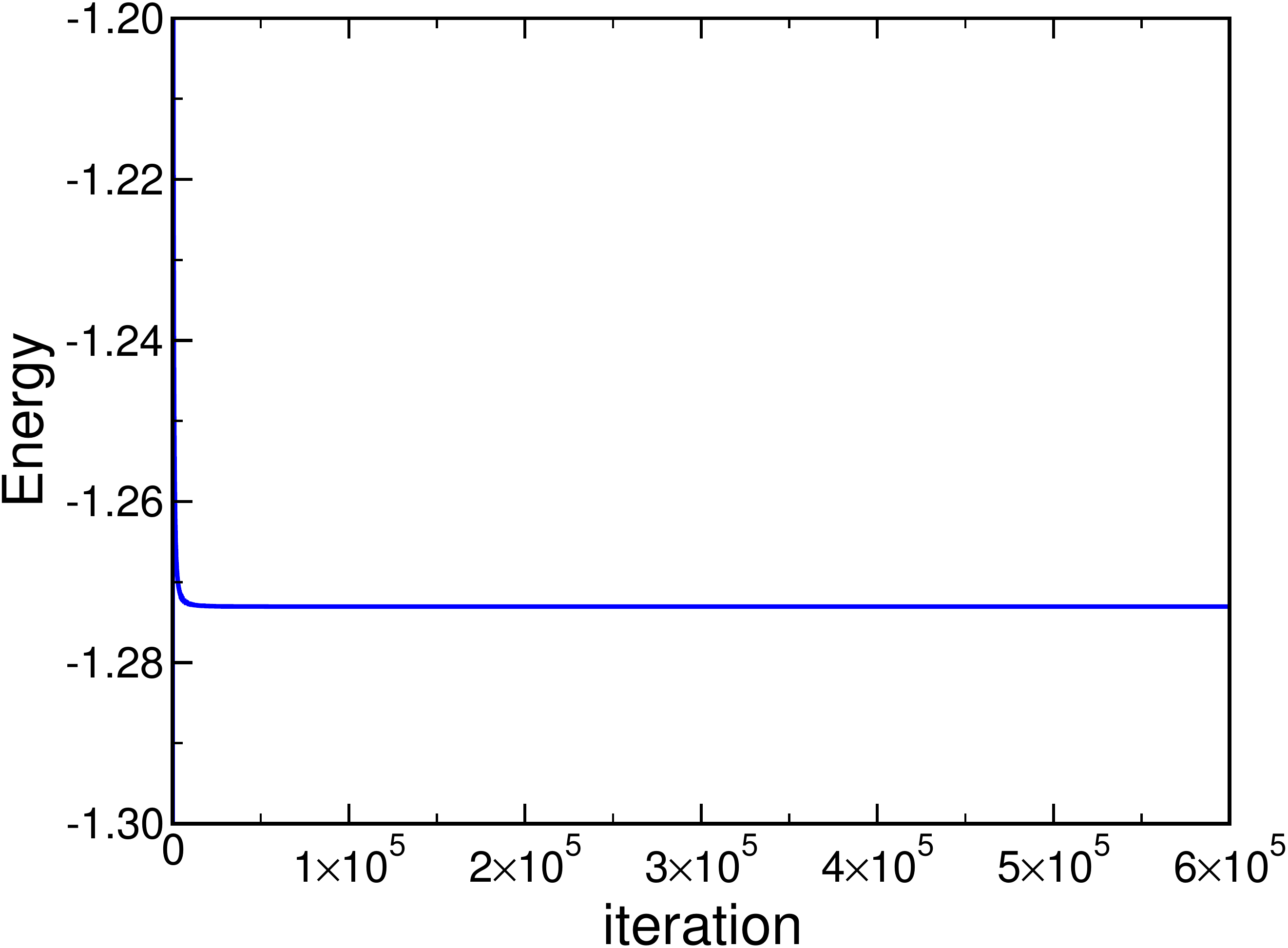}
\vspace*{0.0in}
\caption{(color online)
Variational energy of two-species hard-core Hamiltonian ($H$) as a function of the evolution iteration in iTEBD.
Here, $U=0$, $\mu=0$, and $\epsilon=0.001$.
\label{fig::econv}}
\end{figure}

\begin{figure}[t]
\includegraphics[width=8cm]{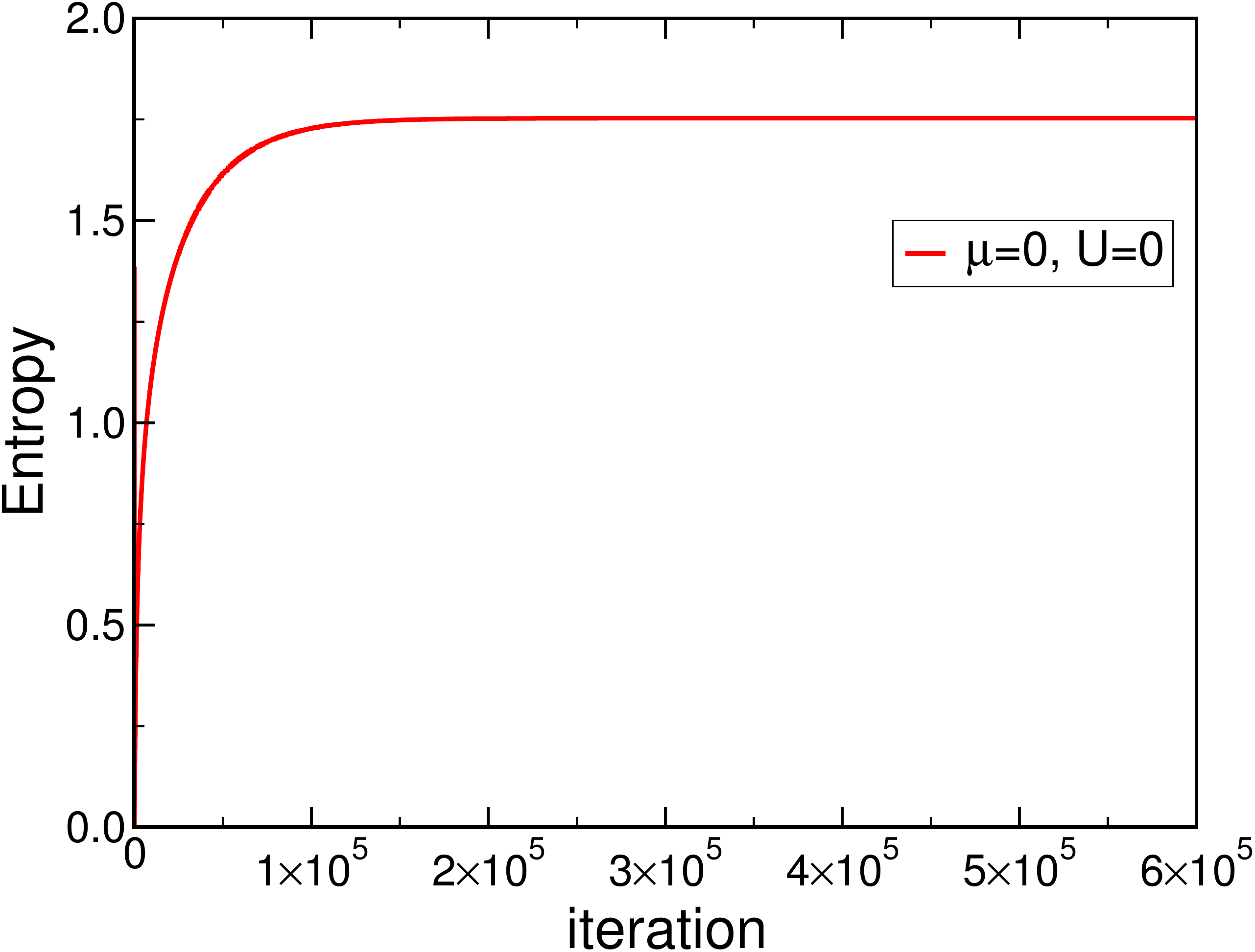}
\vspace*{0.0in}
\caption{(color online)
Entanglement entropy as a function of the evolution iteration in iTEBD.
Here, $U=0$, $\mu=0$, and $\epsilon=0.001$
\label{fig::entconv}}
\end{figure}

Many previous articles have shown that the convergence of energy can be obtained in relatively short iterations of  TEBD\cite{Orus, JWLEE}.
We found that the convergence of other physical quantities should be checked before concluding that we reached reasonable values for them.
Especially, we found that even though the virtual energy converged the exact value, the half-chain entanglement entropy did not reach a reasonable value.
In Figure \ref{fig::econv}, we show the iTEBD energy value versus the number of iterations in TEBD.
In Figure \ref{fig::entconv}, we show the half-chain entanglement entropy versus the number of iterations in iTEBD.
In these two figures, it is clearly shown that the convergence of energy and that of entropy are remarkably different, which we should be very careful in determining EE. 

To overcome this difficulty, we found a new method for fast converging of entropy.
The idea is similar to the temperature annealing method in classical Monte Carlo simulations. 
In iTEBD, we need two things to consider, the first one is that we need $\epsilon$ as small as possible to evolve the system more smoothly from the previous iteration to the next iteration.
The second thing is that, to reach the zero-temperature quantum ground state, we need the number of iteration as large as possible.
It would be nice that we set $\epsilon=0.001$ and let the number of iteration as $2 \times 10^5$ as in Fig. 2.
But the problem is that it takes a long time to run $ 2 \times 10^5 $.
For a fast convergence, we varied $\epsilon$ of time evolution. 
We set $\epsilon=1/4$ during the first $2000$ iterations and reduced $\epsilon$ by multiplying $1/4$ to the previous $\epsilon$ during the next $2000$ iterations.
Because we started from relatively large $\epsilon$, the system evolves rapidly in the beginning, but afterward, it will evolve more slowly.
Figure 3 shows the evolution of entropy as a function of the iteration with varied $\epsilon$.
EE reaches almost the same number as in Fig. 2 with only 12000 iterations, which is crucial in reducing computation time.
The limiting value in Fig. 2 is 1.75298501549855 and the limiting value in Fig. 3 is 1.7542082360011582. 
The difference between the two is $0.07\%$.
We also found that our method increases the accuracy of the ground-state energy more than the accuracy of iDMRG, which will be shown in the next section.

\begin{figure}[t]
\includegraphics[width=8cm]{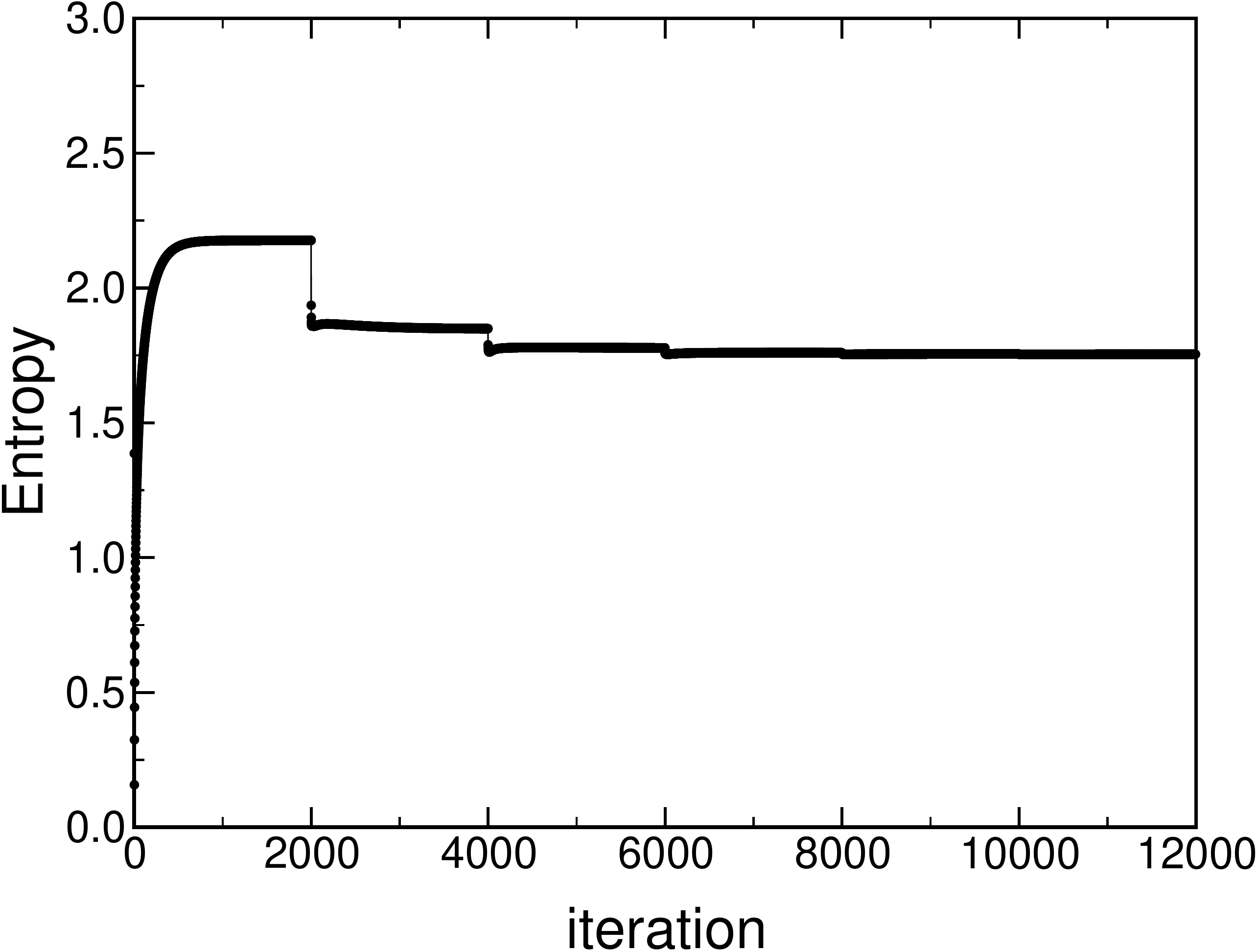}
\vspace*{-0.1in}
\caption{(color online)
Half-chain entanglement entropies (EE) as a function of the evolution iteration in iTEBD.
Starting from $\epsilon=1/4$, at each multiplication of 2000 iterations, we reduced $\epsilon$ by multiplying $1/4$ in time-evolution operator $e^{-\epsilon H}$.  
EE reaches almost the same number as in Fig. 2 with only 12000 iterations, which is crucial in reducing simulation time.
\label{fig::newmethod}}
\end{figure}

\section{Ground-state energies}

The half-filled case is realized when $\mu=0$ in Eq. (\ref{Hamiltonian}).
The only changing parameter is the local interaction ($U$) between two-species bosons.
The exact value of the ground state energy in the thermodynamic limit is well studied by Bethe ansatz by Lieb and Wu \cite{LiebandWu1968, 1DHubbard}.
The ground-state energy per site for the infinite lattice is given by an integral form as
\begin{equation}
E/N = - \frac{U}{4} - 4 \int_0^{\infty} \frac{dx}{x} \frac{J_0(x) J_1(x) }{ \exp( \frac{Ux}{2} ) + 1 }
\label{Bethe_energy}
\end{equation}
for positive $U$ and the energy $E$ for negative $U$ can be obtained from the relation of $E(-U) = E(U)$.

\begin{figure}[t]
\includegraphics[width=8cm]{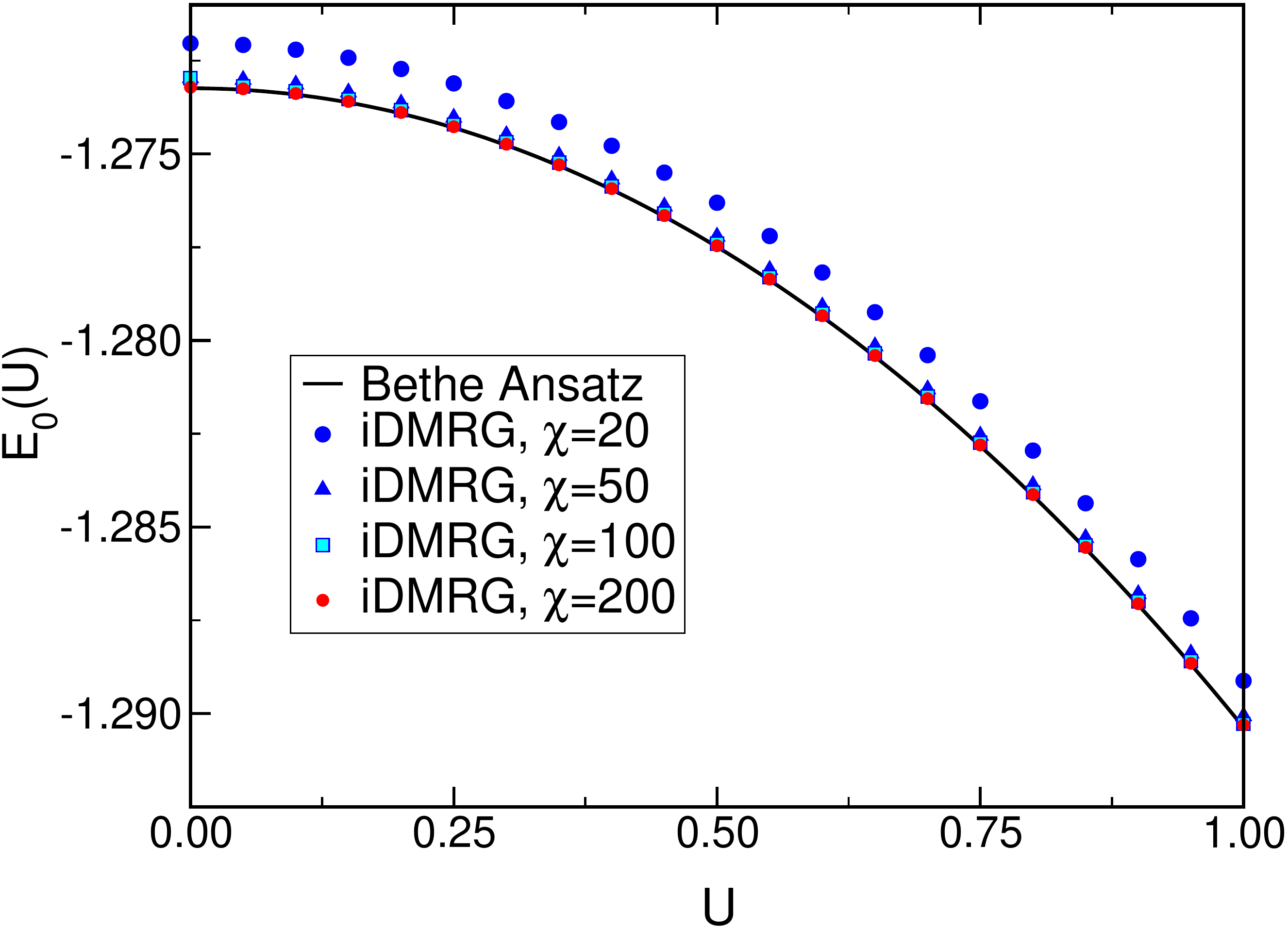}
\vspace*{0.0in}
\caption{(color online)
Variational lowest energies of Eq. (\ref{Hamiltonian}) as a function of $U$ in $[0,1]$ with iDMRG for $\chi=20$, $50$, $100$, and $200$.
For comparison, we plot the Bethe ansatz energy of Eq. (\ref{Bethe_energy}).
The values of iDMRG reaches to the exact ground-state energy with the error of $2.0\times 10^{-5}$ when $\chi=200$.
\label{fig4-ge-dmrg}}
\end{figure}

In Figure \ref{fig4-ge-dmrg}, we show the variational energy for ground states using iDMRG ($\chi$=$20$, $50$, $100$, and $200$) as a function of $U \in [0,1]$, and the reference ground-state energy obtained from Eq. (\ref{Bethe_energy}).
 When $\chi=200$, iDMRG gives fairly good ground-state energies with the typical deviation of $2.0 \times 10^{-5}$ from the true ground-state energy.
Here, we set the criterion of convergence of variational energy as $|E(i+1) -  E(i)| < 10^{-7}$ where $i$ is the iteration index because the typical run of iDMRG generates a monotonic decreasing function of energies versus the iteration.

\begin{figure}[t]
\includegraphics[width=8cm]{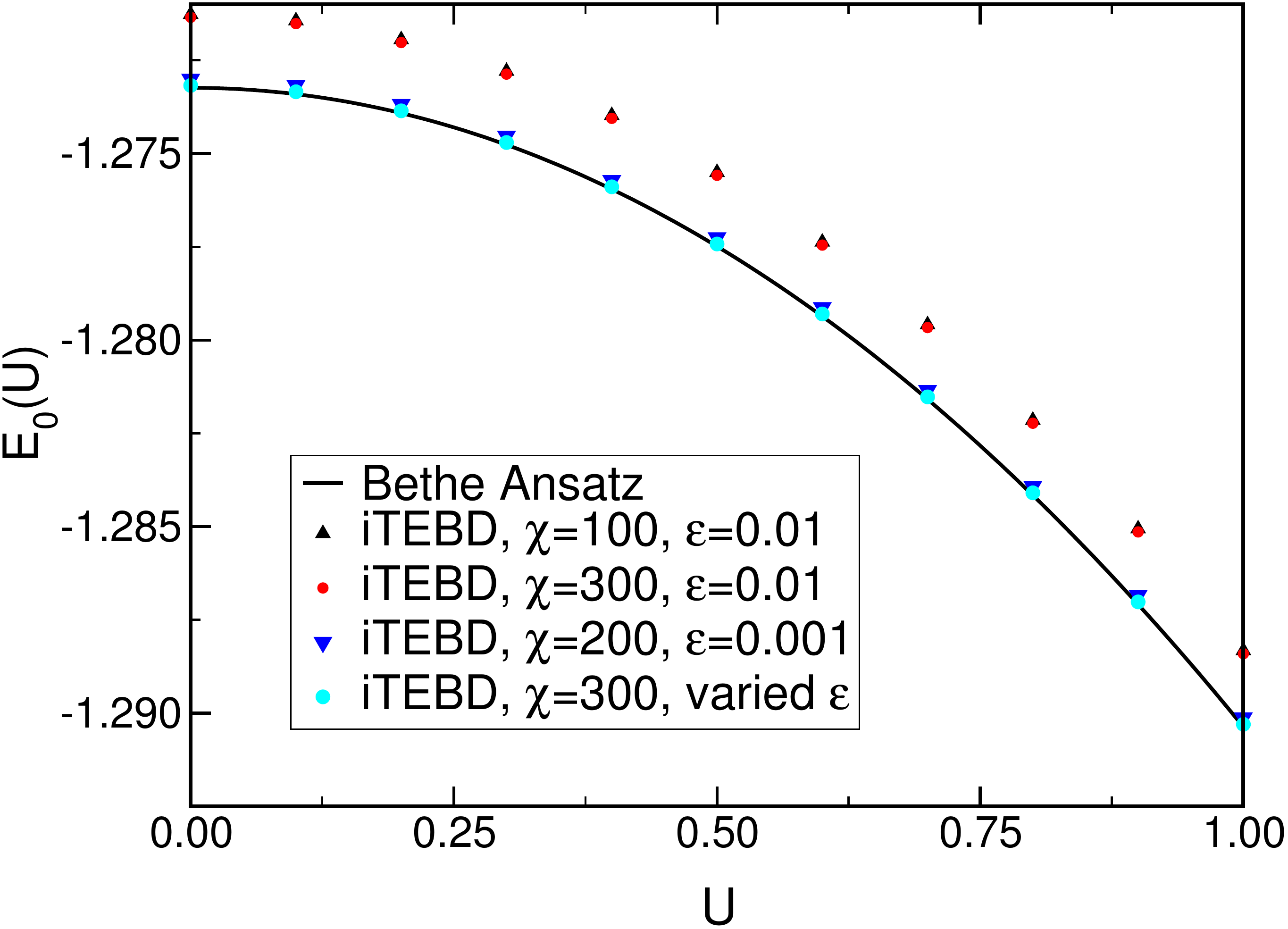}
\vspace*{0.0in}
\caption{(color online)
Variational lowest energies of Eq. (\ref{Hamiltonian}) as a function of $U \in [0,1]$ with various $\chi=100, 200, 300$ and $\epsilon=0.01, 0.001$.
For comparison, we plot the Bethe ansatz energy of Eq. (\ref{Bethe_energy}). 
Here, we note that the smaller $\epsilon$ is more important factor in obtaining the ground-state energy.
\label{fig5-ge-tebd}}
\end{figure}

Figure \ref{fig5-ge-tebd} shows the variational energies for ground states using iTEBD ($\chi$=$100$ and $200$ for $\epsilon$=$0.01$, and $\chi=200$ for $\epsilon$=$0.001$).
Interestingly, the Trotter decomposition parameter $\epsilon$ governs the accuracy of the variational energies, but the overall behaviors are the same.
The typical deviation from the exact ground-state energies are around $2.0 \times 10^{-3}$ for $\epsilon=0.01$ and $2.0 \times 10^{-4}$ for $\epsilon=0.001$.
We believe that we found systematic errors due to the Trotter decomposition and the degree of freedom of singular value decomposition ($\chi$).
This result shows that we approached the very accurate ground-state information from iTEBD and iDMRG.

\begin{figure}[t]
\includegraphics[width=8cm]{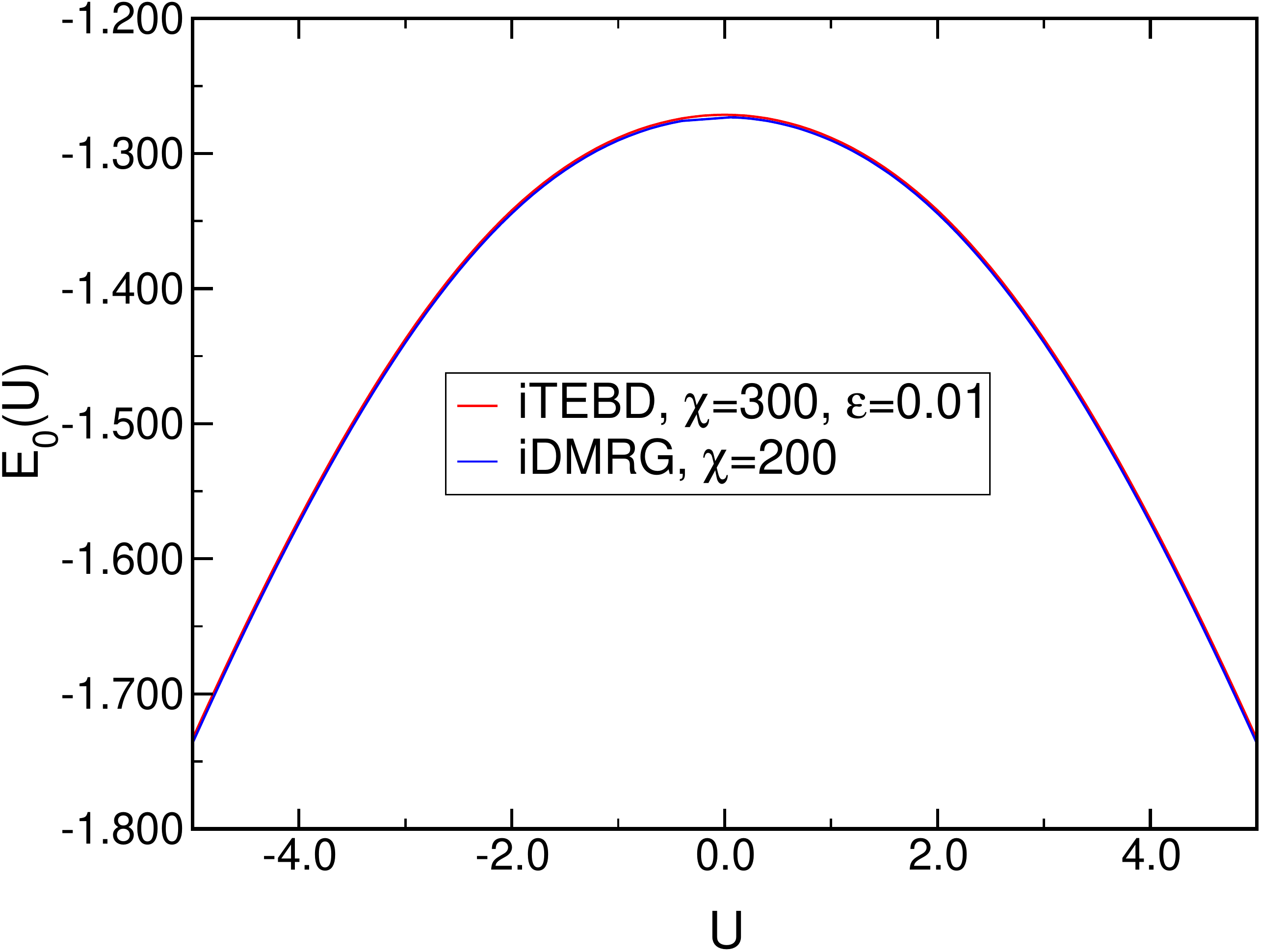}
\vspace*{0.0in}
\caption{(color online)
Variational energies for ground-state energies of two-species hard-core Hamiltonian as a function of $U$ in $[-5,5]$ from iTEBD and iDMRG.
The overall behaviors are the same for both schemes and it is hard to distinguish two in this large energy scale compared with small energy scale in FIG. \ref{fig4-ge-dmrg} and \ref{fig5-ge-tebd}.
\label{fig6-ge-largescale}  }
\end{figure}

Figure \ref{fig6-ge-largescale} shows the overall ground-state energies for $U$ in $[-5,5]$ from iTEBD and iDMRG.
It is hard to distinguish two results because the energy scale is larger than those in Fig. \ref{fig4-ge-dmrg} and \ref{fig5-ge-tebd}.
iTEBD and iDMRG work reasonably well for energy in the broad region of $U$.

\section{Half-chain entanglement entropy}

The half-chain entanglement entropy (von Neumann entropy) is defined from the entanglement of a half-divided chain.
If a quantum state $| \Psi \rangle$ is decomposed with two half chains, $A$ and $B$ by using  a singular value decomposition, $| \Psi \rangle $ can be written as
\begin{equation}
| \Psi \rangle_{\rm SVD} =  \sum_i  \lambda_i | \phi_A  \rangle_i | \phi_B \rangle_i
\end{equation}
where $| \phi_{A(B)} \rangle_i$ is the partial wavefunctions for $A(B)$ and $\lambda_i$ is the singular value.
A normalization condition enforces $\{\lambda \}$ to satisfy
\begin{equation}
1 = {}_{\rm SVD}\langle \Psi | \Psi \rangle_{\rm SVD}   = \sum_i  \lambda_i ^2
\end{equation}
so that the normalized $| \Psi \rangle_{\rm SVD, n} $ is
\begin{equation}
| \Psi \rangle_{ {\rm SVD, n}} =  \sum_i  \frac{\lambda_i}{\sqrt{\sum_j \lambda_j^2} } | \phi_A  \rangle_i | \phi_B \rangle_i.
\end{equation}
The reduced density matrix for the chain $A$ is obtained by tracing out $B$ from full density matrix,
\begin{equation}
\rho_A =  {\rm Tr}_B | \Psi \rangle \langle \Psi |.
\end{equation}
Because $\rho_A$ is diagonal in the space of $| \phi_A \rangle$'s,
The entropy is
\begin{equation}
S_h =  - \sum_i  \frac{\lambda_i^2}{\sum_j \lambda_j^2 } \log \frac{\lambda_i^2}{\sum_j \lambda_j^2 }.
\end{equation}

\begin{figure}[t]
\includegraphics[width=8cm]{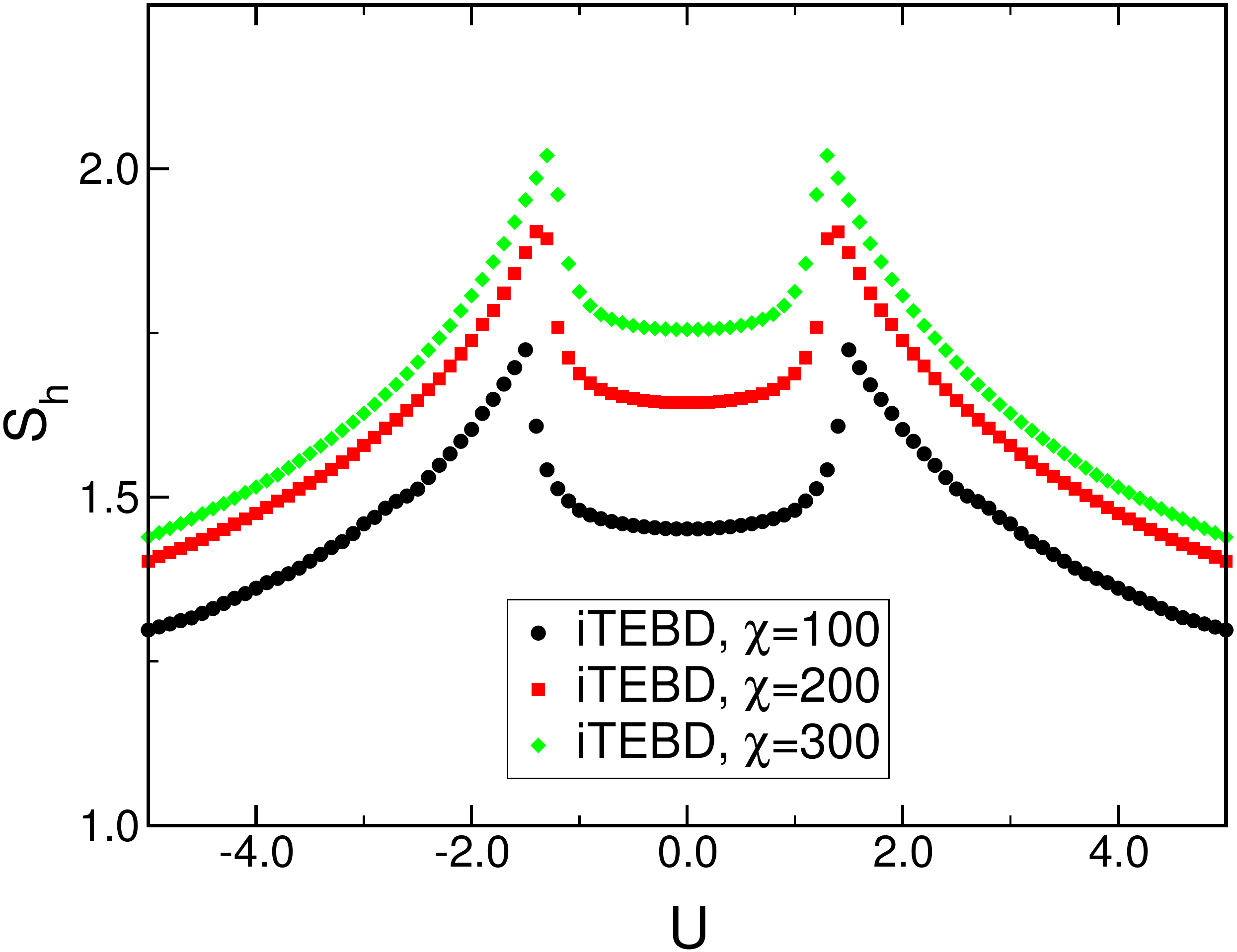}
\vspace*{-0.1in}
\caption{(color online)
Half-chain entanglement entropies ($S_h$) as a function of $U$ from iTEBD.
Note that $S_h$ has a local minimum at $U=0$ and this is the only superfluid phase in half-filled case.
The local maximum points for $S_h$ are located for different $U$, which means that the local maximum point is not related to a quantum critical point.
We claim that one should be careful in identifying a quantum critical point {\it via} half-chain entanglement entropy with its local maximum in other literature.
\label{fig-entropy}}
\end{figure}

With Bethe ansatz, it is very hard to obtain the half-chain entropy in the closed form, but in iMPS, we can easily measure $S_h$ because we already use singular value decompositions for the time evolution of iMPS.
Figure \ref{fig-entropy} shows the half-chain entanglement entropy obtained from iTEBD as a function of $U$ for the half-filled case. 
Because the phase transition occurs at $U=0$, we should focus on the behavior of $S_h$ near $U=0$. The EE has a local minimum at $U=0$ and it increases up to a finite value and decreases as $U$ becomes large.
Also, $S_h$ increases as $\chi$ increases for a fixed $U$. For $U=0$, if $\xi \sim \chi^{\kappa}$, then the fitting form becomes $ \frac{ c \kappa }{6} \log(\chi)$ \cite{Pollmann}. Our fitting result is $c \kappa/6 = 0.2763(5)$, which is consistent with the previous numerical results\cite{Mccha} (0.294) on fermionic Hubbard model with the maximum of $\chi$ was 108.
We also found that $S_h$ does not scale with logarithmic function for $U=5.0$, which might be a characteristic of the Mott insulator phase. 
We do not present $S_h$ form iDMRG because the data is not that clean as iTEBD. 
The issue is connected with the open boundary condition of iDMRG\cite{Chungprivate}.


\section{Off Half-filled case}

\begin{figure}[t]
\includegraphics[width=8cm]{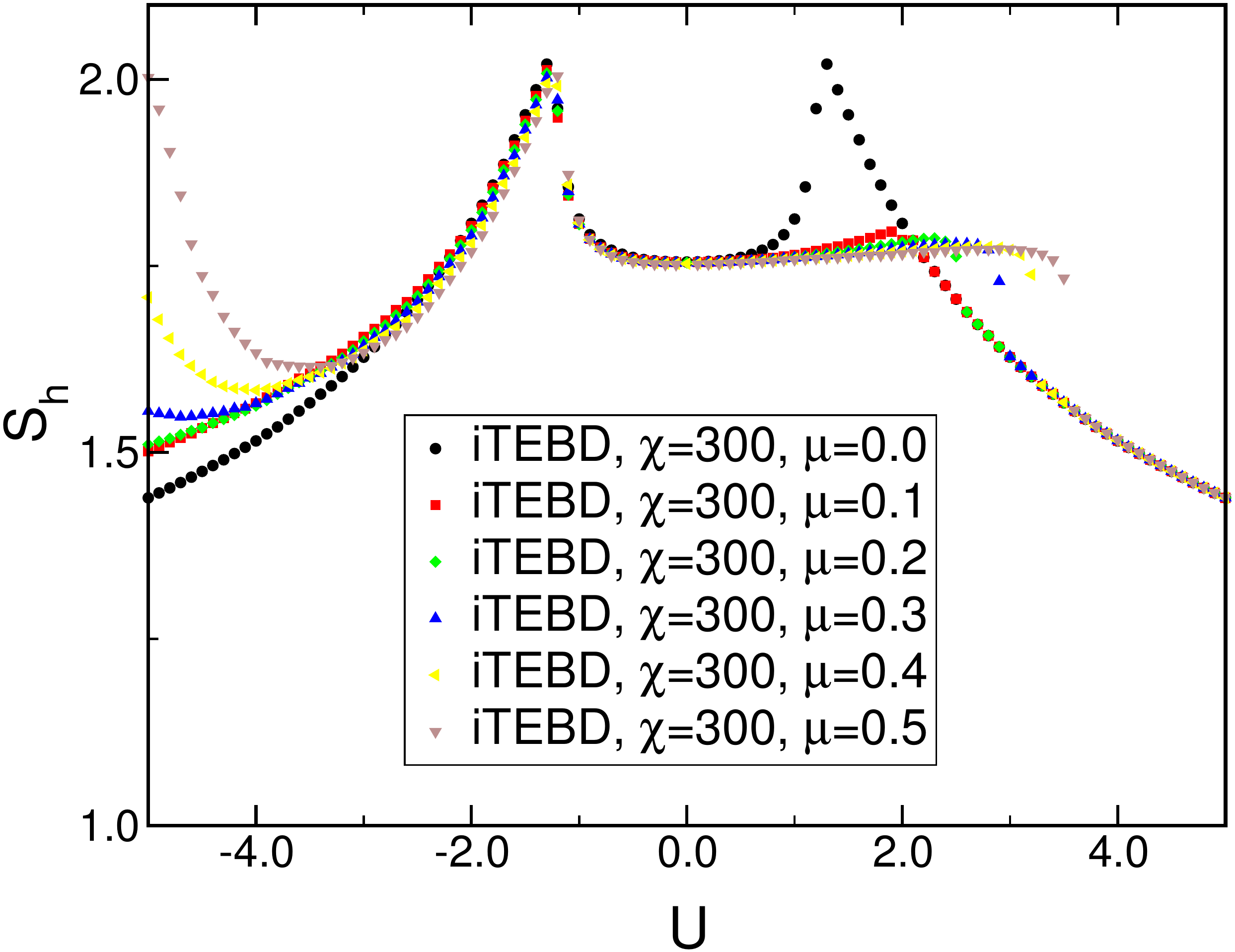}
\vspace*{-0.1in}
\caption{(color online)
Half-chain entanglement entropies ($S_h$) as a function of $U$ from iTEBD for different $\mu$'s ($0.0 \sim 0.5$).
For the density-driven phase transition, EE follows an universal curve when $U>0$.
When $U<0$, as $\mu$ increases, new local minimum points are generated, which means that there is a phase transition where two bosons start making a strong pairing at local sites.
\label{fig-entropyoff}}
\end{figure}

\begin{figure}[t]
\includegraphics[width=8cm]{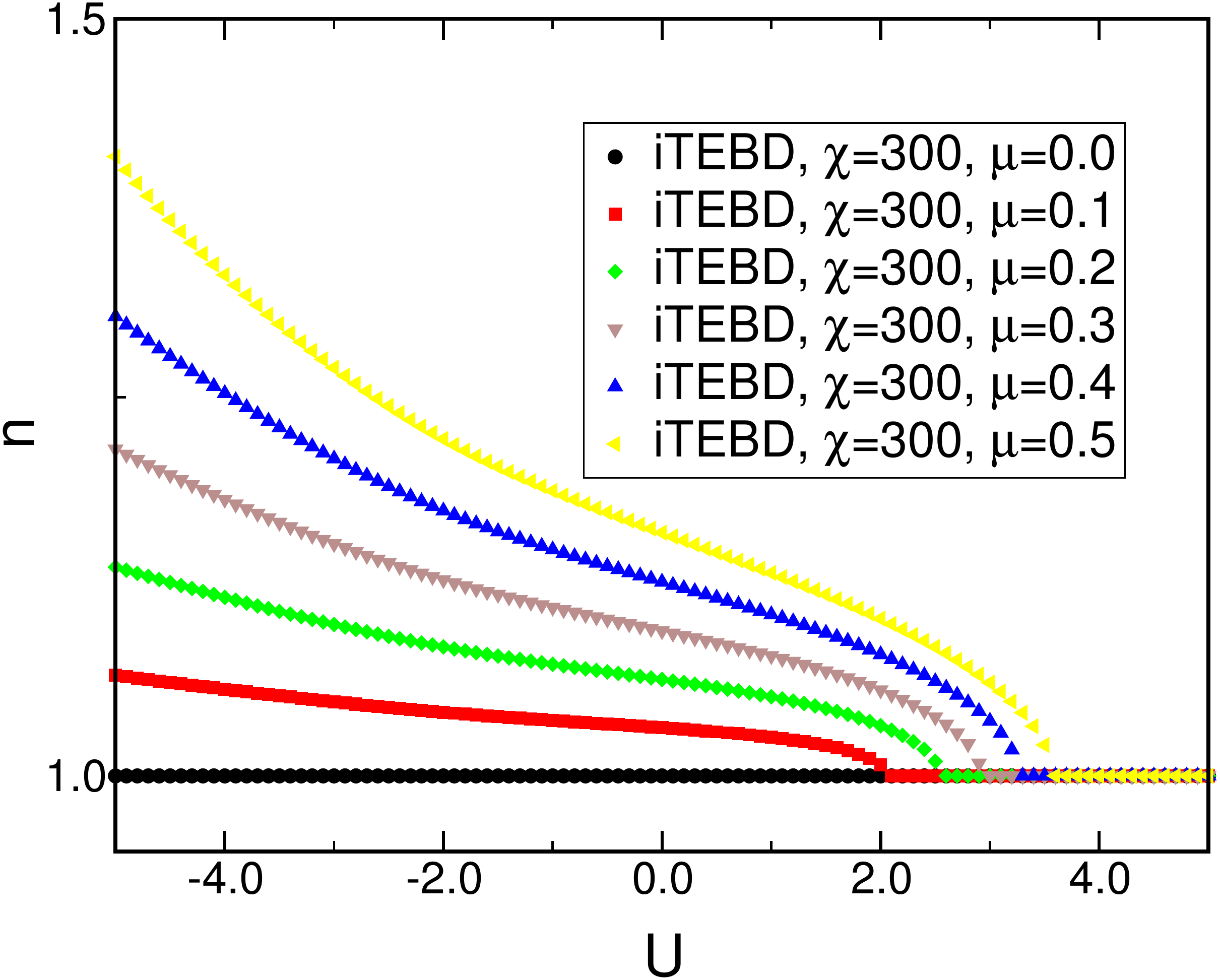}
\vspace*{-0.1in}
\caption{(color online)
Number of bosons ($n_{\rm boson}$) per site as a function of $U$ for different $\mu$'s ($0.1\sim 0.5)$. 
As $U$ increases, the number reduces to 1, which means the system becomes half-filling. 
So, this is a density-driven phase transition, where the EE deviate a universal curve in Fig. \ref{fig-entropyoff}.
\label{fig-nboson}}
\end{figure}

We are also interested in obtaining the entanglement entropy in case of off half-filling.
For nonzero $\mu$, the half-chain entanglement entropy as a function of $U$ is shown in Fig. \ref{fig-entropyoff}.
We obtained the $S_h$ for $\mu=0.1 \sim 0.5$ using iTEBD with $\chi=300$.
Note that $\mu=0.0$ is the half-filled case.
It is quite interesting that the symmetry concerning $U$ is broken here.
To understand the broken symmetry, we measure the density of bosons.

The density of bosons are defined as
\begin{equation}
n_{\rm boson} =  \frac{1}{L} \langle \sum_{i} n_{ia} +  n_{ib}  \rangle
\end{equation}
where $\langle \cdots \rangle$ is the expectation value with respect to a ground state.
Usually, the density is a function of $U$, $\mu$.
In Fig. \ref{fig-nboson}, we fix $\mu$ and plot $n_{\rm boson}$ as a function of $U$.
As $U$ increases, the density reduces from a value above 1 to 1, which means the system becomes again a half-filled Mott insulator.
It is interesting to note that after $U$ reaches a critical value, EE follows a universal curve, and for negative $U$, new local minima appear for $\mu \neq 0 $.
This means that there is a phase transition where different-species bosons make a strong pair at local sites.


\section{Discussion}

In this paper,
we obtained the infinite-size ground-state properties of two-species hard-core bosons in one dimension.
In our model, the bosons have a hard-core condition for the same species, but a local interaction $U$ with the different species when they are at the same site.
In a sense, it is a counterpart of the fermionic Hubbard model in one dimension.
Constructing infinite-size matrix product states for one dimension, we used a time-evolution block decimation method and matrix product operator based on density matrix renormalization group procedure to obtain ground-state energies and half-chain entanglement entropy.
For the half-filling, where the quantum critical point is a single point of $U=0$, we found clear evidence that the half-chain entanglement entropy is a local minimum so that it should be carefully reinterpreted that the maximum entropy point is critical in the previous literature.
Also, we obtained the half-chain entanglement entropies for the off half-filled case.
The symmetry of the entanglement entropy with respect $U$ is broken by the nature of the interaction between the bosons in the upper Hubbard band and if $U$ is large enough, then the Mott insulator with boson density of $1$ prevails again.

We also used the original DMRG algorithm by S. White\cite{White} and obtained the entanglement entropy with density matrices. 
For large $U$, the entanglement oscillates with iterations.
This is due to the open boundary condition \cite{Chungprivate}.
The same phenomena occurs with iDMRG. 

We are working on the soft-core boson models which will demonstrate more interesting features of bosons in optical lattices and the half-chain entanglement of the model will give shed light to understand the properties of the quantum ground state of bosons.

\acknowledgments
This research was supported by Basic Science Research Program through the National Research Foundation of Korea(NRF) funded by the Ministry of Education(NRF-2017R1D1A1B03035188).
JWL thanks the Korea Institute for Advanced Study (KIAS) for its hospitality under the Associate Member program.
JWL also thanks M.-C. Cha, M. H. Chung, T. Xiang, and Z. Yuan for their stimulating discussions.

\newpage

\end{document}